# Towards Probability-based Safety Verification of Systems with Components from Machine Learning

Hermann Kaindl
*Institute of Computer Technology*
TU Wien
Vienna, Austria
kaindl@ict.tuwien.ac.at

Stefan Kramer
*Institute of Computer Science*
*Johannes Gutenberg-Universität Mainz*
Mainz, Germany
kramer@informatik.uni-mainz.de

*Abstract*—Machine learning (ML) has recently created many new success stories. Hence, there is a strong motivation to use ML technology in *software-intensive systems*, including *safety-critical* systems. This raises the issue of *safety verification of ML-based systems*, which is currently thought to be infeasible or, at least, very hard. We think that it requires taking into account specific properties of ML technology such as: (i) Most ML approaches are *inductive*, which is both their power and their source of error. (ii) Neural networks (NN) resulting from deep learning are at the current state of the art not transparent. Consequently, there will always be errors remaining and, at least for deep NNs (DNNs), verification of their internal structure is extremely hard. In general, safety engineering cannot provide full guarantees that no harm will ever occur. That is why *probabilities* are used, e.g., for specifying a *risk* or a *Tolerable Hazard Rate* (THR).

In this vision paper, we propose verification based on probabilities of errors both estimated by controlled experiments and output by the inductively learned classifier itself. *Generalization error bounds* may propagate to the probabilities of a hazard, which must not exceed a THR. As a result, the quantitatively determined bound on the probability of a classification error of an ML component in a safety-critical system contributes in a well-defined way to the latter's overall safety verification.

## I. INTRODUCTION

Software-intensive systems, in particular cyber-physical systems (CPS), need to have essential properties like security and safety. While security especially of deep neural networks (DNN) has already been studied theoretically [1], there is an urgent need for a new safety framework. Hence, the focus of this paper is on safety verification of systems with components from machine learning (but not on systems that learn while being applied).

In practice, safety is taken very seriously, e.g., for railway systems, see EN 50126, and there is a relatively new automotive standard for functional safety (of traditional cars), ISO 26262. It does not address safety assurance of ML-based functions, however. This was a key motivation for ISO/PAS 21448 on safety of the intended functionality (SOTIF), where Annex I (informatively) mentions a few of the issues involved in using ML in safety-critical systems. The Automated Driving Roadmap ERTRAC 17 [2] contains the challenge of applying artificial intelligence, in particular (deep) ML, with regard to safety. There are no hints, however, on how to address it.

According to the standard ISO/IEC Guide 51:2014, a safety risk combines the severity of harm with the probability that it occurs. Also THR and (A)SIL – (Automotive) Safety Integrity Level – are defined based on probabilities. Hence, we approach the difficult issue of safety verification of ML-based software-intensive systems on the basis of probabilities.

In fact, we propose to be even more ambitious than the current approach to verification of conventional automotive software according to the standard ISO 26262, where no explicit estimation of probabilities is included. The safety assurance principle of ISO 26262 assumes that an adequate level of rigor can reduce the residual risk of a hazard due to software failure to an acceptable level. And according to IEC 61508-3, quantification of probabilities is not required.

In this paper, we propose a new theoretical framework for safety verification of software-intensive systems including ML technology. It is based on existing work on safety verification, but has to take into account that especially for DNN only black-box approaches will be applicable in the near future, since they are currently considered opaque. Including controlled experiments for such a purpose is new for practical use, and the way of managing the data for such experiments will be new as well. In particular, this new framework involves determining a probabilistic generalization error bound, i.e., an upper bound on the true generalization error based on a quantity that is calculated on the training set, e.g., the empirical error plus some term depending on a complexity measure. One typical form of bound would be that, for any $\delta > 0$, it holds with probability at least $1-\delta$ that for all hypotheses or models $h \in H$: $Error_{rtue}(h) \leq Error_{emp}(h) + f(H, D, \delta)$. where $D$ is the training set. One central component of generalization bounds are complexity measures.

An essential problem to be solved is that presenting instances for learning requires taking into account the right context in the application and in the final model, such that specific instances or sets of instances become relevant whenever needed. This has to be taken into account in the determination of hazard rates or risks, too.

The remainder of this paper is organized in the following manner. First, we sketch some background material and the state of the art, in order to make this paper self-contained. After that, we lay out a new framework for safety verification. Focusing on one of its key parts, we then sketch how one can take advantage of recent progress with probabilistic generalization error bounds for deep learning for the purpose of safety engineering. Finally, we draw a tentative conclusion and propose future work.

## II. BACKGROUND AND STATE OF THE ART

For the *safety integrity* of preexisting *software*, IEC 61508-7-D is an *informative* appendix of this safety standard. It provides guidance for an approach to estimation of probabilities from statistics gained from previous use of exactly this software in the course of operations. It may be used, for instance, to show that a piece of software originally built according to the requirements of SIL1 of IEC 61508-3, also fits SIL2, after an appropriate time of use and an appropriate number of applications during successful operations.

Recently, Salay and Czarnecki [3] adapted and extended the standard ISO 26262 to address safety assurance of ML-based functions, making a detailed assessment and adaption of ISO 26262 for ML, specifically in the context of supervised learning. They carefully analyzed and adapted each requirement in this standard and, where gaps were found, they proposed additional requirements. Still, they kept the *safety assurance principle* of ISO 26262 that developing software using an adequate level of rigor can reduce the residual risk of hazard due to software *failure* (termination of its ability to perform a function as required) to an acceptable level. We propose to be more ambitious through experimentally determined probabilities, aiming for higher guarantees of safety assurance for DNN than currently for traditional software that is yet to be put into operation. We consider this very important for potential medical applications as well.

Annex I of ISO/PAS 21448 essentially mentions unintended bias or distortion in the collected data as problems, referring to [4]. In fact, one of the most common assumptions in ML is that the training set and test sets are drawn from the same underlying distribution (*i.i.d.* – independently drawn from the identical distribution), that is, the training situation resembles the test situation. Of course, it is important that the training data actually corresponds to the intended operational domain. Without any further precautions or mechanisms, only then and with an increasing number of instances, learning, in the sense of getting measurably better at a task with increasing evidence, can take place.

*Verification of DNN models* is currently viewed as very hard, due to their high nonlinearity emerging from a high number of layers, and the variety of different architectures and topologies [5]. This work and Huang et al. [6] make a claim about safety verification while, in fact, both deal with a kind of robustness, more specifically against adversarial perturbations. It was taken up for confidence claims for analyzing robustness in [7], where more generally *confidence arguments* are presented for evidence of performance in machine learning for highly automated driving functions. We think that such arguments could be stronger if based on probabilities.

In [8], the problem of accidents in machine learning systems is discussed by taking particular properties of ML into account, but not even mentioning *induction*. Generally, inductive machine learning is a case of *inductive generalization*, which is by definition not truth-preserving and an unsafe type of inference [9]. That is why it is common to study generalization on the basis of *learning curves*, where the *x*-axis represents the number of training instances and the *y*-axis represents a chosen performance measure (for a recent example in the context of software product lines, see [10]).

If probabilistic guarantees for the performance of an ML model are required, one can resort to *generalization error bounds* [11, 12]. Considering DNN, it has been observed that their behavior is sometimes unstable, i.e., their output varies largely depending on only small perturbations of the input. First approaches have addressed this problem from a technical point of view [13], e.g., by changing the training protocol or by alternative loss functions. From a higher point of view, the problems and variants are not much different than laid out already long time ago [14]. Nevertheless, it has been observed that error bounds based on traditional computational learning theory (e.g., based on the VC dimension or the Rademacher penalty) do not work for deep learning models, because of their large capacity. This is in contradiction to their often outstanding generalization performance. Recent progress in this area [15, 16] does not only improve this situation, but promises to be applicable in practice, at least in combination with empirical error measurement. For a discussion of the relevant issues and topics, we refer to a recent survey paper [17].

A more recent, large-scale experimental comparison shows the correlation of current generalization bounds and their underlying complexity measures with actual generalization performance [18]. The results suggest that many of the existing margin- and norm-based bounds fail at "predicting" the actual generalization performance. However, the results also suggest that bounds and generalization measures centered around the notion of "sharpness" of local minima show great promise in that respect. In particular, magnitude-based perturbation bounds (see Section 4.3.1 there and Appendix C.3 for actual calculations) exhibit the highest mutual information with generalization overall.

## III. THE NEW FRAMEWORK FOR SAFETY VERIFICATION

In the new framework, we consider the overall approach as an iterative, systematic and defined process. In particular, we propose to iterate over the following steps:

1. Data preparation and learning (as usual);
2. Estimation of probabilities of *errors* (wrong results from the inductively learned classifier) through controlled experiments and determination of a probabilistic generalization error bound;
3. If this error bound is insufficient, repeat by going back to 1 with additional data, or otherwise, exit successfully.

As long as the outcome in terms of the generalization error bound is considered insufficient, inductive learning will have to be continued with additional training data, most importantly also including "edge cases" or "near misses". From a statistical perspective, this poses the problem of the usual assumption that the distribution for testing is identical to the distribution for training. Changing the training procedure in the way proposed above may invalidate this assumption. Hence, some care will have to be taken to ensure it. And again, the training data need to correspond to the intended operational domain. It should be noted that the proposed procedure makes heuristic use of statistics and statistical learning theory, but due to its theoretical underpinning it is still more reliable than working solely with empirical errors.

Practically useful generalization bounds for deep learning are just about to emerge. A recent paper (and its updates) based on compression [16] improves upon previous approaches (e.g., [15]), and hints at practical applicability (see Fig. 4 in that paper, for a comparison with existing bounds and between the bound and the measured error). While the scale of the bound still differs by magnitudes, the rankings of the models according to the empirical error and according to the bound are correlated. A very recent, large-scale empirical comparison [18] (see also above) showed that so-called "sharpness-based measures", e.g., magnitude-aware perturbation bounds, have the highest current overall correlation with actual generalization performance. This suggests that such a bound could already be used, perhaps together with other measures, for *model selection*, i.e., picking a suitable model from a set of several, e.g., differently parameterized candidate models.

For making such an approach sound, also the inclusion of context for the instances involved is relevant. Developing a general scheme for the applicability domain of ML models will contribute to more predictable behavior. The applicability domain has been a concept for machine learning models only in chemical risk assessment so far [19], but could prove valuable in more general settings. The idea is that models should be allowed to interpolate, but not to extrapolate. The applicability domain defines the scope or competence of a classifier. Within the applicability domain, the error rates, in particular the false-positive and false-negative rates, are demanded to remain below specified thresholds. *Abstaining classifiers* (i.e., classifiers with the reject option) are also likely to play a role, because there will be instances outside the applicability domain. It will be possible to choose machine learning models with desired properties dynamically and systematically according to the current context.

Finally, the results of generalization error bounds along with the empirical error can be propagated to the hazard or risk analysis of the overall safety-critical system, see below for a sketch.

IV. HOW TO USE PROBABILISTIC GENERALIZATION ERROR BOUNDS

At the current state of the art, usually *functional safety engineering* is applied to safety-critical systems development, which involves the identification of safety-related functions. We are interested in cases where a classifier resulting from ML delivers such a function. In such a case, it will also have to be included in probabilistic hazard or risk assessment of this system. Once the causality of some hazard as related to the ML component is understood, the contribution of its error probability to the probability of function failure, and indirectly to the probability of this hazard and related risks of the overall system can be determined. More precisely, we propose to use a probabilistic generalization error bound of the ML component for these analyses, since this will increase safety as compared to directly using error probabilities (see above for the difference).

Actually, a probability of *failure* will have to be determined, based on both the measured empirical error and a probabilistic generalization error bound of the ML component. While an error just means a wrong answer, a failure means a complete termination of the ability of the ML component to perform its function as required. For example, if an ML component will be deployed on a dedicated computer system possibly even including special-purpose hardware (like a DNN), its *failure* may also be caused by a failure of this computer system, and this has to be taken into account additionally.

For hazard or risk analyses, the envisaged use of the safety-related functions and their possible failure must be analyzed, in order to identify possible *hazards* in the context of this particular system. For an overview of the key concepts involved, see Fig. 1 taken from [20], where we studied these concepts and their relationships qualitatively both for automotive and railway safety standards. First, we created a *core ontology of safety risk concepts*, since the terminology and even the concepts of the automotive standard ISO 26262 are not fully aligned with the ones for railway [21]. Fig. 1 additionally includes the concept of a *risk*. Since safety *risks* involve both the *severity* and the *probability* of some harm possibly caused by the failure of a classifier resulting from ML, determining its probabilistic generalization error bound is key.

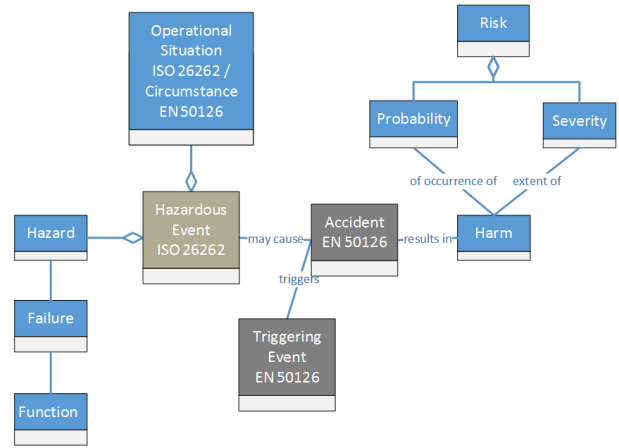

Fig. 1. Conceptual model defining a core ontology of safety risk concepts (diagram taken from [20]).

As shown in Fig. 1, the operational situation or circumstance is relevant as well, and this will even reinforce the problem of *applicability domains* of the classifier resulting from ML. We conjecture that it may be handled analogously to hardware, which has different probabilities of failure, e.g., at different temperatures.

V. CONCLUSION AND FUTURE WORK

For (random) failures of hardware components, assessing their probabilities and using them for hazard or risk analyses is common practice for a long time now. For software failures especially of new software, in contrast, there is no practical approach for really determining their probabilities at the current state of the art (where IEC 61508-7-D presents a statistical approach based on previous use of exactly the same software, i.e., without any single change). For failures caused by errors of inductively learned classifiers, however, we claim that our approach as outlined here has the potential to provide generalization error bounds in the future.

This vision paper brings together that for classifiers resulting from ML, generalization error bounds may be provided, and that these, together with empirically estimated error probabilities, can then be used for risk or hazard analyses in a mathematically founded way. The prospective benefits of such an approach are high, since there maybe even 'better' safety cases for ML-based systems in practice than the current ones for traditional software, where usually no probability calculations are done for determining a SIL or ASIL. Hence, the overall scientific relevance for the field of Information and

Communication Technology is very high, in order to make ML technology applicable within software-intensive systems, in particular safety-critical systems.

Future work will have to develop a mechanism that recognizes changing distributions, outliers and edge cases, and explicitly handles them dynamically in a statistically sound way. Different distributions may be modeled for different contexts, with an additional layer that recognizes the context and picks the right distribution or part of a model [22]. An example in the domain of driving would be one distribution or model for driving on a Californian highway, and another one for an old Italian city. Different contexts would require different distributions or parts of models. Generally speaking, presenting hand-picked instances for learning requires taking into account in the application and in the final model, that the right context is picked, such that specific instances or sets of instances become relevant whenever needed.

After the availability of these theoretical results as well as their take-up for engineering practice, both the societal and the economic prospects are great. The safety risks of ML-based systems will be precisely known, and only 'sufficiently' safe systems may be used, leading to less harm and greater economic value.

.